\documentclass[article,twocolumn,showpacs,superscriptaddress,preprintnumbers,floats,aps,pra,10pt]{revtex4-1}
\usepackage{graphicx}
\usepackage{dcolumn}
\usepackage{amsmath}
\usepackage{amsfonts}
\usepackage{amssymb}
\usepackage{color}
\usepackage{float}
\usepackage{pstricks}
\usepackage{natbib}
\usepackage{array}
\usepackage{multirow}
\usepackage{ulem}
\providecommand{\U}[1]{\protect\rule{.1in}{.1in}}
\providecommand{\abs}[1]{\lvert#1\rvert}

\DeclareMathOperator\arctanh{arctanh}

\begin{document}
\title{Singularities and Soft-Big Bang in a viscous $\Lambda$CDM model}
\author{Norman Cruz}
\altaffiliation{norman.cruz@usach.cl}
\affiliation{Departamento de F\'isica, Universidad de Santiago de Chile, \\
Avenida Ecuador 3493, Santiago, Chile}
\affiliation{Center for Interdisciplinary Research in Astrophysics and Space Exploration (CIRAS), Universidad de Santiago de Chile, Avenida Libertador Bernardo O'Higgins 3363, Estación Central, Chile}
\author{Esteban Gonz\'alez}
\altaffiliation{esteban.gonzalez@uac.cl}
\affiliation{Direcci\'on de Investigaci\'on y Postgrado, Universidad de Aconcagua, Pedro de Villagra 2265, Vitacura, 7630367 Santiago, Chile}
\author{Jose Jovel}
\altaffiliation{jose.jovel@usach.cl}
\affiliation{Departamento de F\'isica, Universidad de Santiago de Chile, \\
Avenida Ecuador 3493, Santiago, Chile}
\date{\today}

\begin{abstract}
\begin{center}
\textbf{{Abstract:}}    
\end{center}
In this paper we explore the different types of singularities that arise in the $\Lambda$CDM model when dissipative processes are considered, in the framework of the Eckart's theory. In particular, we study the late-time behavior of $\Lambda$CDM model with viscous cold dark matter (CDM) and an early-time viscous radiation domination era with cosmological constant (CC). The fluids are described by the barotropic equation of state (EoS) $p=(\gamma-1)\rho$, where $p$ is the equilibrium pressure of the fluid, $\rho$ their energy density, and $\gamma$ is the barotropic index. We explore two particular cases for the bulk viscosity $\xi$, a constant bulk viscosity $\xi=\xi_{0}$, and a bulk viscosity proportional to the energy density of the fluid $\xi=\xi_{0}\rho$. Due to some previous investigations that have explored to describe the behavior of the universe with a negative CC, we extend our analysis to this case. We found that future singularities like Big-Rip are allowed but without having a phantom EoS associated to the DE fluid. Big-Crunch singularities also appears when a negative CC is present, but also de Sitter and even Big-Rip types are allowed due to the negative pressure of the viscosity, which opens the possibility of an accelerated expansion in AdS cosmologies. We also discuss a very particular solution without Big Bang singularity that arises in the early-time radiation dominant era of our model known as Soft-Big Bang.
\vspace{0.5cm}
\end{abstract}

 \maketitle


\section{Introduction}
It is well known in current cosmology that the accelerated expansion of the universe is one of the most fascinating puzzles in physics. This behavior is supported by the cosmological data coming from measurements of Supernovae type Ia (SNe Ia) \cite{Riess1998,sPerlmutter,Planck2018}, the observational Hubble parameter data (OHD) \cite{OHD2021}, the baryonic acoustic oscillations (BAO) \cite{BAO2017}, the cosmic microwave background (CMB) \cite{M.Teg}, and information from large-scale structures (LSS) formation coming from WMAP \cite{WMAP2013}; showing also that the universe is spatially flat. 

There are different approaches in order to describe this accelerated expansion of the universe. One of them is to add in the energy-momentum tensor $T_{\mu \nu}$, in the right hand side of the Einstein gravity equation, an exotic fluid with negative pressure, dubbed dark energy (DE), which can cause an overall repulsive behavior of the gravity at large cosmological scales (see \cite{L.S.Universe,D.MandD.E,D.E} for some excellent reviews). The other approach is by modifying the left hand side of the  Einstein equation, i.e., the geometry of the space time, that leads to different ideas of modify gravity (for some theories of modified gravity which involve this idea see \cite{introduccionMG,gravedadmoficiada5,referee1,referee2}). For the first approach, the most simple model is the $\Lambda$CDM model which it is also the best cosmological model in order to describe the cosmological data \cite{Planck2018,WMAP2013}. In this model, the current universe is dominated by dark matter (DM) and DE, representing approximately the 30\% and 70\% of the total energy density of the universe, respectively. This DE is given by the CC ($\Lambda$), which can be characterized by a barotropic EoS with barotropic index $\gamma=0$, causing the acceleration in the universe expansion \cite{Riess1998}. However, this model is not absents of problems, of which we can highlight:
\begin{itemize}
    \item The CC problem: the value of the CC predicted from field theoretical estimations is about 60-120 order of magnitude larger than the observed value \cite{Lambda,CPWeinberg,problemadecoincidenciaylambda}.
    \item The coincidence problem: current energy densities of DM and DE have the same order of magnitude, but in the $\Lambda$CDM model, these energy densities evolves differently, so, it is necessary a fine-tuning between them in the early universe in order to that both densities match in order of magnitude at the current time \cite{C.Problem,referee,problemadecoincidenciaylambda}.
    \item The $H_{0}$ tension: measurements of the Hubble parameter at the current time $H_{0}$ shows a discrepancy of $4.4\sigma$ between the value inferred from Planck CMB and the locally meassurements obtained by A. G. Riess \textit{et al.} \cite{Riess:2019cxk}.
    \item EDGES: most recently results of the experiment to detect the global EoR signature (EDGES) detect an excess of radiation that is not predicted by the $\Lambda$CDM model in the reionization epoch, specifically at z$\approx 17$ \cite{linia21cm}.
\end{itemize}

One approach to overcome some of these problems, without going further than $\Lambda$CDM or modify the gravity, is to consider disipative fluids as a more realistic way of treating cosmic fluids \cite{bulkyaceleracion,impactofviscosity,remedyforplankanlssdata}. In this sense, several authors shown that a bulk viscous DM in different models without DE can cause the accelerated expansion of the universe \cite{paperprofeAccelerated,Bulk1,Bulk2,Bulk3,Bulk4,Bulk5,Bulk6,Bulk7,Bulk8,Exact,Testing}, due to the negativeness of the viscous pressure, which allows to alleviate in principle the CC and the coincidence problems. The excess of radiation predicted by EDGES are explain in \cite{linea21cmH}, where the authors consider a viscous nature in DM. In \cite{tensionH0,BulktensionH} the authors address the $H_{0}$ tension problem as a good chance to construct new cosmological models with viscous/inhomogeneous fluids in the context of a bulk viscosity. Furthermore, tensions in the measurements of $\sigma_{8}-\Omega_{m}$ (where $\sigma_{8}$ is the r.m.s. fluctuations of perturbations at $8h^{-1}Mpc$ scale) and $H_{0}-\Omega_{m}$ coming from LSS observations and the extrapolated from Planck CMB parameters using the $\Lambda$CDM model, can be alleviate if one assumes a small amount of viscosity in the DM \cite{remedyforplankanlssdata}. Some authors as also used bulk viscosity in inflationary phases of the universe \cite{earlyandlateviscous,otrainflacionbulk}. 

It is important to mention that from Landau and Lifshitz \cite{LandauandLifshitz} we know that the bulk viscosity in the cosmic evolution seems to be significant and we can interpreted from the macroscopic point of view that is equivalent to the existence of slow processes of restoring equilibrium state. Some authors proposal that bulk viscosity of the cosmic fluid may be the result of non-conserving particle interactions \cite{Cosmologycreation} and another have shown that different cooling rates of the components of the cosmic medium can produce bulk viscosity \cite{cooling1,cooling2,cooling3}. Also, for neutralino CDM bulk viscosity pressure appears in the CDM fluid due to the energy transferred from the CDM fluid to the radiation fluid \cite{hofmann2001damping}. Many observational properties of disk galaxies can be reproduced by a dissipative DM component, which appears as a residing component in a hidden sector \cite{foot2015dissipative,foot2016solving}. On the other hand, at perturbative level viscous fluid dynamics provides a simple and accurate framework for extending the description into the nonlinear regime \cite{blas2015large}. Since up to date it is unknown the nature of the DM and the dissipative effect in cosmology can not be discarded, it is of physical interest to explore the behavior of this type of DM in the $\Lambda$CDM model.

In order to study dissipative processes in cosmology it is necessary to development a relativistic thermodynamic theory out of equilibrium, being Eckart the first who developed it \cite{Eckart}. Later, it was discovered that Eckart's theory was not really relativistic, since it is a non-causal theory \cite{NocausalEckart,Muller}. A causal theory was proposed by Israel and Stewart \cite{I.S.1,I.S.2}, but it presents a much greater mathematical difficulty than the Eckart's theory, even in scenarios where the bulk viscosity doesn't present very exotic forms. Therefore, many authors work in the Eckart's formalism in order to have a first approximation of the cosmological behavior with dissipative fluids \cite{Big.Bang,AlmadaEckartbulktest,bulkyaceleracion,chinos,articuloprofephantom,Brevik}, since the Israel-Stewart's theory is reduced to the Eckart's theory if the relaxation time for transient viscous effects is equal to zero. \cite{Dissipativecosmology}

As we mentioned before, in both Eckart's and Israel-Stewart's theory it is possible to describe the accelerated expansion of the universe without the inclusion of a CC. Nevertheless, as it was previously discussed by Maartens \cite{Dissipativecosmology}, in the context of dissipative inflation, the condition to have an accelerated expansion due only to the negativeness of the viscous pressure enters into direct contradiction with the near equilibrium condition that is assumed in the Eckart's and Israel-Stewart's theory
\begin{equation}\label{equilibrio}
    \left|\frac{\Pi}{p}\right|\ll 1,
\end{equation}
which means that the viscous stress $\Pi$ must be lower than the equilibrium pressure $p$ of the dissipative fluid. So, following this line, it as been prove in \cite{Analysing,articulobueno} that the inclusion of a positive CC could preserve the near equilibrium condition (\ref{equilibrio}) in some regime. The price to pay is to abandon the idea of unified DM models with dissipation as models that can consistently describe the late time evolution of the universe. It is important to mention that a negative CC can not be ruled out from study in cosmology \cite{constantenegativa,transicionlambda,negativa1,negativa2,negativa3,negativa4,negativa5,negativa6,negativa7}. For example, a negative CC appears naturally in superstring theory in the dual space $AdS_{5}\times S^{5}$  \cite{st3,st2,st1}. Some authors even mentioned the possibility of a transition between a negative CC to a positive one \cite{transicionlambda,turco}. Even more, a negative CC has been explored by many authors with the aim of alleviating the $H_{0}$ tension \cite{negativa1,negativa2,negativa3,negativa4,negativa5,negativa6,negativa7}. 

Works with dissipation where the CC is included have been studied in recent times, for example the authors in \cite{chinos} already work in Eckart Formalism with CC and a bulk viscosity proportional to the Hubble parameter, or more interesting scenarios can be seen in \cite{fullsolution} where the authors also include a CC that is variable in time. On the other hand, some authors have shown that the presence of bulk viscosity in the DE could cause that their effective barotropic index can be less than $0$ \cite{articuloprofephantom,Brevik}. Fluids with a barotropic index $\gamma<0$ are dubbed ``phantom" \cite{phantommenace} and can not be ruled out of the current cosmological data. For example, some works indicated that the barotropic index of the DE is inconsistent with the value of $0$ at $2.3\sigma$ level \cite{WMAP2013,2sigmalevel}. The possibility of phantom EoS for the DE open an interesting scenario known as Big-Rip, in which the scale factor presents a singularity in a finite future time \cite{BigSmash(B.R)}. Following this line, the authors in \cite{tiposdeBigRip,clasificacionBigRipdetallada} have made a classification of the different singularities obtained in models with phantom DE. 

Other interesting issue related to viscous fluid is the possibility of avoid singularities. In the framework of general relativity many studies found cosmological scenarios where there are no singularities,  corresponding to emergent and bouncing universes \cite{Generalized,linking,universoemergente,Softbang1,SoftBang2}. A regular universe without Big Bang was found in \cite{Big.Bang}, where the viscosity drives the early universe to a phase with a finite space-time curvature. This regular scenario called ``soft Big-Bang" are also discussed in other contexts \cite{Softbang1,SoftBang2}, describing universes with eternal physical past time, that come from a static universe with a radius greater than the Planck radius to be far out of the regime where quantum gravity has to be employed.

The aim of this paper is to explore exact solutions of a viscous $\Lambda$CDM like model, looking for the conditions that leads to early and late time singularities, considering a bulk viscosity term constant and proportional to the energy density of the dissipative fluid. These two simple cases open up a great variety of behaviors which will allow us to study different singularity scenarios in the framework of the Eckart's theory. We will discuss our results according to the classification given in \cite{tiposdeBigRip}. Also, we will investigate solutions which represent regular universes, as it was found by Murphy in \cite{Big.Bang}, but with the inclusion of a CC. It is important to note that many authors have studied these type of singularities within the framework of cosmological models filled with a phantom DE \cite{Brevik,brevik2,tiposdeBigRip,D.E,clasificacionBigRipdetallada}. In our case, the model can be characterized by an effective EoS that represent the behavior of the two fluids of the model, the dissipative fluid and the CC, as a whole. Even more, we will study solutions where the CC can take negative values. Therefore, in this work we will try to give a more complete understanding of the early and late-time singularities when dissipative process are considered in a $\Lambda$CDM like cosmological model.

The outline of this paper is as follow: In section II we describe briefly the non-causal Eckart's theory and we find the general differential equation to solve. In section III we present the possibility of de Sitter like solutions that arises from the general differential equation previously found. In section IV we start by describing briefly the different types of singularity that arises for a Friedmann–Lemaître–Robertson–Walker (FLRW) metric. In subsection A we study the late-time singularities that arises in our model for a constant dissipation, an a dissipation proportional to the energy density of the dissipative fluid, for a positive CC. In subsection B we will do the same for the case of a negative CC. In subsection C we discuss early-time singularities for the case of positive CC. In subsection D we discuss early-time singularities for the case of negative CC. In section V we discuss an early-time solution without Big Bang singularity called ``soft Big-Bang". Finally, in section VI we present some conclusions and final discussions. $8\pi G=c=1$ units will be used in this work.

\section{Theory of Eckart with CC}

In what follows, we will consider a flat FLRW cosmological spacetime, dominated by only two matter components: a DE given by $\Lambda$, and a barotropic fluid with EoS $p=(\gamma-1)\rho$, where $p$ is the equilibrium pressure of the fluid, $\rho$ their energy density and $\gamma$ is the barotropic index, that takes the values of $\gamma=1$ for CDM and $\gamma=4/3$ for radiation. This barotropic fluid experience dissipative processes during their cosmic evolution, with a bulk viscosity coefficient $\xi$ that depends on their energy density through the power-law
\begin{equation}\label{xilaw}
\xi=\xi_{0}\rho^{m}, \,\,\,\, \xi_{0}>0,
\end{equation}
where $\xi_{0}$ and $m$ are constant parameters, with $\xi_{0}>0$ in order to be consistent with the second law of thermodynamics \cite{librocaro}. The behavior described by Eq. (\ref{xilaw}) for the viscosity has been widely investigated in the literature as one of the simplest and most natural choices since the bulk viscosity of fluids depends, particularly, in its temperature and pressure, and therefore it is physically suitable to take this dependence. Other elections include, for example, the function $\xi=\xi_{0}+\xi_{1}H$ \cite{chinos}, but in this case and since we are including a CC, this election implies that the viscosity of the fluid is a function not only of its properties but also of the CC.  

In the Eckart's theory, the field equations in presence of bulk viscous are
\begin{equation}\label{tt}
H^2=\Big(\frac{\dot{a}}{a}\Big)^{2} =\frac{\rho}{3}+\frac{\Lambda}{3},
\end{equation}
\begin{equation}\label{rr}
\frac{\ddot{a}}{a}=\dot{H}+H^2=-\frac{1}{6} \left(\rho+3P_{eff}\right)+\frac{\Lambda}{3},
\end{equation} 
where ``dot'' accounts for the derivative with respect to the cosmic time $t$, $a$ is the scale factor, $H$ the Hubble parameter, and $P_{eff}$ is an effective pressure given by
\begin{equation}\label{Peff}
P_{eff}=p+\Pi,
\end{equation}
being $\Pi$ the bulk viscous pressure defined in the Eckart's theory by
\begin{equation}\label{Pi}
\Pi=-3H\xi.
\end{equation}
The conservation equation takes the form
\begin{equation}\label{ConsEq}
\dot{\rho}+3H(\rho+p+\Pi)=0.
\end{equation}
Therefore, we can obtain from Eqs.~(\ref{xilaw})-(\ref{ConsEq}) a single evolution equation for $H$, which is given by
\begin{equation}\label{Hpunto}
2\dot{H}+3\gamma H^{2}-3\xi_{0}H(3H^{2}-\Lambda)^{m}-\Lambda\gamma=0.
\end{equation}

Since we are interested in comparing some solutions of Eq. (\ref{Hpunto}) for different values of $m$ with the standard $\Lambda$CDM model, we display below the solution for $H(t)$ y $a(t)$ with the initial conditions  $H(t=0)=H_{0}$ and $a(t=0)=1$, for the case without dissipation ($\xi=0$)
\begin{equation}\label{Hestandar}
H(t)=\frac{H_{0}\sqrt{\Omega_{\Lambda}}\left(\left(\sqrt{\Omega_{\Lambda} }+1\right) e^{3 \gamma  H_{0} t \sqrt{\Omega_{\Lambda} }}-\sqrt{\Omega_{\Lambda} }+1\right)}{\left(\sqrt{\Omega_{\Lambda} }+1\right) e^{3\gamma H_{0} t \sqrt{\Omega_{\Lambda} }}+\sqrt{\Omega_{\Lambda} }-1},  
\end{equation}
\begin{equation}\label{aestantdar}
    a(t)=\left(\cosh\left(\frac{3\gamma\sqrt{\Omega_{\Lambda}}H_{0}t}{2}\right)+\frac{\sinh\left(\frac{3\gamma\sqrt{\Omega_{\Lambda}}H_{0}t}{2}\right)}{\sqrt{\Omega_{\Lambda}}}\right)^{\frac{2}{3\gamma}},
\end{equation}
where $\Omega_{\Lambda}=\Lambda/(3H^{2}_{0})$.  From Eq. (\ref{Hestandar}) we can see that $H=\sqrt{\Lambda/3}$ for very late times, corresponding to the de Sitter behavior.

\section{de Sitter like solutions}

Before to make a complete integration of Eq. (\ref{Hpunto}), we will explore the possibility of de Sitter like solutions. Knowing this behavior will help us to compare with the asymptotic behaviors in cases when $\dot{H}\neq0$. Taking $H=H_{dS}$ with $\dot{H_{dS}}=0$, Eq. (\ref{Hpunto}) reduces to the following algebraic equation
\begin{equation}\label{Halgebraic}
3\gamma H_{dS}^{2}-3\xi_{0}H_{dS}(3H_{dS}^{2}-\Lambda)^{m}-\Lambda\gamma= 0.
\end{equation}

One general result can be quickly found if the above equation is written as
\begin{equation}\label{Halgebraicmodif}
(3H_{dS}^{2}-\Lambda) \left [\gamma-3\xi_{0}H_{dS}(3H_{dS}^{2}-\Lambda)^{m-1}\right]=0,
\end{equation}
which indicates that the values of $H_{dS}$ given by
\begin{equation}\label{HdeSitter}
H_{dS}=\pm \sqrt{\frac{\Lambda}{3}},
\end{equation}
are two real solutions of Eq. (\ref{Halgebraicmodif}), for $m\geq1$ and $\Lambda>0$. Note that the positive solution corresponds to the usual de Sitter one and the contracting solution $H_{dS}<0$ it is not of physical interest. The other possible de Sitter solutions are obtained taking the square bracket of the left-hand side of Eq. (\ref{Halgebraicmodif}) equal to zero, for different values of $m$.

\subsection{Case $m=0$}
In this case, the dissipation of the fluid is constant and Eq. (\ref{Halgebraicmodif}) becomes in a quadratic equation of the form
\begin{equation}\label{Halgebraic0}
H_{dS}^{2}-\frac{\xi_{0}}{\gamma} H_{dS}-\frac{\Lambda}{3}=0,
\end{equation}
with a discriminant given by
\begin{equation}\label{discriminant0}
\Delta_{0} =\left(\frac{\xi_{0}}{\gamma}\right)^{2}+\frac{4\Lambda}{3}.
\end{equation}
Then, two solutions are allowed for the Hubble parameter
\begin{equation}\label{Hsolutionpositive}
H_{dS\pm}=\frac{\left(\xi_{0}/\gamma\right)\pm\sqrt{\Delta_{0}}}{2}.
\end{equation}

The above equation depends on the values of $\xi_{0}$, $\gamma$ and $\Lambda$, and three type of solutions are obtained depending if $\Delta_{0}$ is positive, zero or negative. This last one, where $\Lambda<-3\xi_{0}^{2}/4\gamma^{2}$, is discarded because represents a complex Hubble parameter without physical interest. If $\Delta_{0}=0$, the solution reduces to
\begin{equation}\label{Hsolutiondeltacero}
    H_{dS}=\frac{\xi_{0}}{2\gamma}\;\; \textup{for}\;\; \Lambda=-\frac{3\xi_{0}^{2}}{4\gamma^{2}},
\end{equation}
being the only de Sitter like solution of the model for this case,  which is driven by the dissipative processes. Since $\xi_{0}$ can be expressed in terms of $|\Lambda|$ it is straightforward to find that in this case $H_{dS}$ in Eq. (\ref{Hsolutiondeltacero}) can also be expressed as $H_{dS}= \sqrt{\frac{|\Lambda|}{3}}$. If $\Delta_{0}>0$, then $\Lambda>-3\xi_{0}^{2}/4\gamma^{2}$, and the model for this case has only two de Sitter like solutions, $H_{dS+}$ and $H_{dS-}$, again directly driven by the dissipative processes. But, $H_{dS-}$ only represent an expanding solution when $\Lambda<0$.

On the other hand, using the Eq. \eqref{tt} it is possible to obtain the energy density associated to the de Sitter solutions \eqref{Hsolutionpositive} and \eqref{Hsolutiondeltacero}, given respectively by
\begin{equation} \label{rhodeSitterpm0}
    \rho_{\pm}=\frac{3\xi_{0}}{2\gamma}\left(\frac{\xi_{0}}{\gamma}\pm\sqrt{\Delta_{0}}\right),
\end{equation}
\begin{equation}\label{rhodeSitter0}
    \rho=\frac{3\xi_{0}^{2}}{2\gamma^{2}}.
\end{equation}
From the above expressions it is possible to see that $\rho_{+}>0$ and $\rho>0$, i. e., the de Sitter like solution given by Eqs. \eqref{Hsolutionpositive} (positive one) and \eqref{Hsolutiondeltacero} do not have null fluid energy density, contrary to the usual de Sitter solution \eqref{HdeSitter} (positive one), where $\rho_{dS}=0$ (DE dominant solution). It is important to note that $\rho_{-}>0$ leads to the constraint $\Lambda< 0$, expression that it's according with the constraint obtained in order to $H_{dS-}$ represent an expanding solution. Therefore, the de Sitter like solutions with physical interest for $m=0$ are $H_{dS\pm}$ and $H_{dS}$.

\subsection{case m=1} \label{seccionsitterm1}
In this case the dissipation is proportional to the energy density of the dissipative fluid, and the other real solution of Eq. \eqref{Halgebraicmodif}, besides the positive de Sitter solution \eqref{HdeSitter} when $\Lambda>0$, is given by 
\begin{equation}\label{H1deSitter}
H_{dS}=\frac{\gamma}{3\xi_{0}},
\end{equation}
which depends only of the values of $\gamma$ and $\xi_{0}$, i. e., being a de Sitter like solution that is a function of the parameters related to the dissipative processes and, in principle, independent of the values of $\Lambda$. But, using Eq. \eqref{tt}, we obtain that the fluid energy density for this solution is given by
\begin{equation}\label{rhodeSitter1}
    \rho=\frac{\gamma^{2}}{3\xi_{0}^{2}}-\Lambda,
\end{equation}
expression that when we impose $\rho>0$ leads to $\Lambda<\gamma^{2}/3\xi_{0}^{2}$. Again, this de Sitter like solution do not have null energy density, contrary to the usual de Sitter solution, except when $\Lambda=\gamma^{2}/3\xi_{0}^{2}$. The same result given by Eq. \eqref{H1deSitter} was found in \cite{Big.Bang} for the case of a null CC.

A surprising results in both, $m=0$ and $m=1$ cases, is the possibility of de Sitter like solutions of physical interest despite the presence of a negative CC. It will find that the corresponding exact solutions behaves asymptotically with the de Sitter like evolution found in this section. 

\section{Singularities in viscous $\Lambda$CDM models}

In what follows we will study the solutions that arise from Eq. (\ref{Hpunto}), for the particular cases when $m=0$ and $m=1$, and we discuss their behavior in terms of the free parameters $\xi_{0}$, $\gamma$ and $\Lambda$. The solutions for each case will be compared with the $\Lambda$CDM model. We will focus our study in the existence of different types of early and late time singularities, which can occur for some values of the free parameters of each model, following the classifications given in \cite{tiposdeBigRip,clasificacionBigRipdetallada}:

\begin{itemize}
\item \textbf{Type 0A}  (``Big Bang"): for $t\rightarrow 0$ , $a \rightarrow 0$, $\rho \rightarrow \infty$ and $|p|\rightarrow \infty$. \\
 \item \textbf{Type 0B} (``Big Crunch"): for $t\rightarrow t_{s}$ , $a \rightarrow 0$, $\rho \rightarrow \infty$ and $|p|\rightarrow \infty$. \\
\item\textbf{Type I}  (``Big Rip"): for $t\rightarrow t_{s}$, $a \rightarrow \infty$, $\rho \rightarrow \infty$ and $|p|\rightarrow \infty$. \\
\item\textbf{Type $\mathbf{I_{l}}$}  (``Little-Rip"): for $t\rightarrow \infty$, $a \rightarrow \infty$, $\rho \rightarrow \infty$ and $|p|\rightarrow \infty$. \\
\item \textbf{Type II}  (``Sudden"): for $t\rightarrow t_{s}$, $a \rightarrow a_{s}$, $\rho \rightarrow \rho_{s}$ and $|p|\rightarrow \infty$. \\
\item \textbf{Type III} (``Big freeze") : for $t\rightarrow t_{s}$, $a \rightarrow a_{s}$, $\rho \rightarrow \infty$ and $|p|\rightarrow \infty$. \\
\item \textbf{Type IV} (``Generalized Sudden"):  for $t\rightarrow t_{s}$, $a \rightarrow a_{s}$, $\rho \rightarrow 0$ and $|p|\rightarrow 0$, and higher derivatives of $H$ diverge.
\end{itemize}
These singularities are typical in the following cosmological scenarios: (i) type I emerges at late times in phantom DE dominated universes \cite{doomsday,BigSmash(B.R),phantom1,phantom2,phantom3,phantom4}; (ii) type II corresponds to a sudden future singularity \cite{tipoII,tiposdeBigRip}, also know as a big brake or a big démarrage, which appear under the conditions  $\rho> 0$ and $\rho + 3p >0$ (SEC) in an expanding universe \cite{tipoII2}; type III occurs for models with $p=-\rho-A\rho^{\alpha}$ and the difference with the Big-rip type I is that here the scale factor has a finite value in a finite time \cite{tiposdeBigRip,tipoIII}; and (iv) type IV which also appears in the context of phantom DE of the form $p=-\rho-f(\rho)$, explored in \cite{tiposdeBigRip} with a particular form of $f(\rho)$ called ``32'', and in the context of quantum cosmology \cite{tipoIVdos}.

Since the singularities are characterized by the divergences in the curvature scalar, we will use in our study the Ricci scalar, given by the following expression 
\begin{equation}\label{ricci}
    R=6\left({\frac {{\ddot {a}}(t)}{a(t)}}+{\frac {{\dot {a}}^{2}(t)}{a^{2}(t)}}\right)= 6\left(\dot{H}+2H^{2}\right),
\end{equation}
in order to explore the divergences in the solutions found.

\subsection{Late-times singularities with $\Lambda>0$}

In this subsection we will study the singularities that arise from the solutions of Eq. \eqref{Hpunto} for a positive CC when $m=0$ and $m=1$. In order to compare with $\Lambda$CDM model, we will set in the general solutions $\gamma=1$ (CDM) and $\Omega_{\Lambda}=0.69$, which is the current value given by the cosmological data \cite{Planck2018}. From now, all solutions will be expressed in terms of the dimensionless density parameters $\Omega_{\Lambda}$ and  $\Omega_{\xi}=3^{m}\xi_{0}H^{2m-1}_{0}$, using the initial conditions $H(t=0)= H_{0}$ and $a(t=0)=1$, where $t=0$ is the present time.

\subsubsection{Cases for $m=0$}\label{prueba}
The integration of Eq. (\ref{Hpunto}) is straightforward and leads to an integral of the form $\int\frac{dH}{R}=-\frac{3\gamma}{2}t+C$, where $R=H^{2}-(\xi_{0}/\gamma)H-(\Lambda/3)$ is a polynomial in H. In principle, three different types of solutions emerge depending if the discriminant $\Delta_{0}$, given by Eq. \eqref{discriminant0}, is positive, negative or zero. For $\Lambda>0$ the only solution is with $\Delta_{0}>0$. The condition (\ref{discriminant0}) in terms of  dimensionless densities takes the form $\Delta_{0}=H^{2}_{0}\bar{\Delta}_{0}$, where
\begin{eqnarray}\label{OmegaLambda}
\bar{\Delta}_{0}=\left(\frac{\Omega_{\xi}}{\gamma}\right)^{2}+4\Omega_{\Lambda}>0,
\end{eqnarray}
and $\Omega_{\xi}=\xi_{0}/H_{0}$. The exact solution for this case is
\begin{equation}\label{H1}
    E(T)=\frac{\sqrt{\bar{\Delta}_{0}}}{2}\tanh{\left[\frac{3\gamma\sqrt{\bar{\Delta}_{0}}T}{4}+\arctanh{\left(\frac{2-\frac{\Omega_{\xi}}{\gamma}}{\sqrt{\bar{\Delta_{0}}}}\right)} \right]}+\frac{\Omega_{\xi}}{2\gamma},
\end{equation}
\begin{equation}\label{factordescala1mcero}
    a(T)=\exp{\left(\frac{\Omega_{\xi}}{2\gamma}T\right)}\left\{\frac{\cosh{\left[\frac{3\gamma\sqrt{\bar{\Delta}_{0}}}{4}T+\arctanh{\left(\frac{2-\frac{\Omega_{\xi}}{\gamma}}{\sqrt{{\bar{\Delta}}_{0}}}\right)} \right]}}{\cosh\left[\arctanh{\left(\frac{2-\frac{\Omega_{\xi}}{\gamma}}{\sqrt{\bar{\Delta}_{0}}}\right)}\right]}\right\}^{\frac{2}{3\gamma}},
\end{equation}
where $E(T)=H(T)/H_{0}$ and $T=H_{0}t$ is a dimensionless time, therefore their positive values represents future evolution. It is important to note that the Hubble parameter \eqref{H1} doesn't exhibit a singularity for any time $T$ and the scale factor \eqref{factordescala1mcero} represent a bouncing universe. Even more, the asymptotic behavior of the Hubble parameter for $T\rightarrow\infty$ give us $H_{ds+}$ and for $T\rightarrow-\infty$ gives us $H_{ds-}$, both solutions given by Eq. \eqref{Hsolutionpositive}, being $H_{dS+}$ the de Sitter-like solution of this model.  

\subsubsection{Case for  $m=1$}\label{seccionm1}

\begin{figure}[H]
\centering
\includegraphics[width=0.5\textwidth]{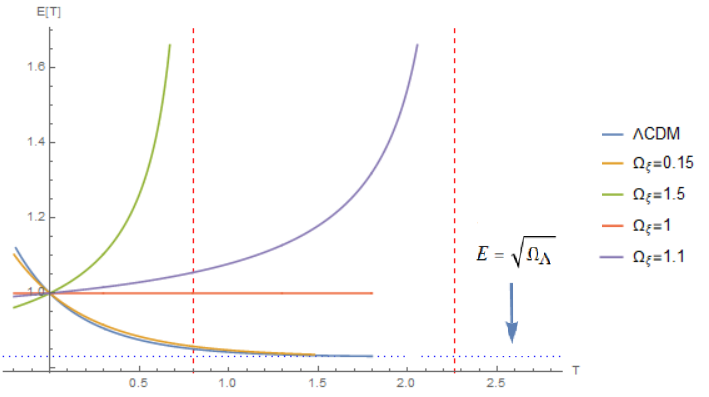}
\caption{{Numerical} behavior of $E(T)$, given by Eq. \eqref{Tm1}, for different values of $\Omega_{\xi}$ and for the particular values of $\gamma=1$ and $\Omega_{\Lambda}=0.69$. We also plotted the $\Lambda$CDM model. The red dashed lines represent the times of singularities given by Eq. \eqref{TiempoBR} for $\Omega_{\xi}=1.5$ and $\Omega_{\xi}=1.1$, respectively.}
\label{graficoEdeT}
\end{figure}

In this case the polynomial in $H$ is $R=(1-3\xi_{0}H/\gamma)(H^{2}-\Lambda/3)$ and the solution takes the dimensionless form

\begin{eqnarray}\label{Tm1}
&T(E)=\frac{\Omega_{\xi} \sqrt{\Omega_{\Lambda}} \log \left(\frac{(1-\Omega_{\Lambda} ) (\gamma - E\Omega_{\xi} )^2}{\left(E^2-\Omega_{\Lambda} \right) (\gamma - \Omega_{\xi} )^2}\right)}{{3 \sqrt{\Omega_{\Lambda} } \left(\gamma^2-\Omega^{2}_{\xi} \Omega_{\Lambda} \right)}} \nonumber \\
& +\frac{\gamma  \log \left(\frac{\left(\sqrt{\Omega_{\Lambda} }-1\right) \left(\sqrt{\Omega_{\Lambda} }+E\right)}{\left(\sqrt{\Omega_{\Lambda} }+1\right) \left(\sqrt{\Omega_{\Lambda} }-E\right)}\right)}{{3 \sqrt{\Omega_{\Lambda} } \left(\gamma^2-\Omega^{2}_{\xi} \Omega_{\Lambda} \right)}},
\end{eqnarray}
where $\Omega_{\xi}=3\xi_{0} H_{0}$. In Fig. \ref{graficoEdeT} we have numerically found the behavior of $E(T)$ given by the above equation. Note that $T\rightarrow \infty, \;\forall E$ when $\Omega_{\xi}=\gamma$, in other words, this case represents the de Sitter case given by Eq. \eqref{H1deSitter}, that is $H(t)=H_{0}$, $\forall t$, as it can be seen from Fig. \ref{graficoEdeT}.

\begin{figure}[H]
\centering
\includegraphics[width=0.5\textwidth]{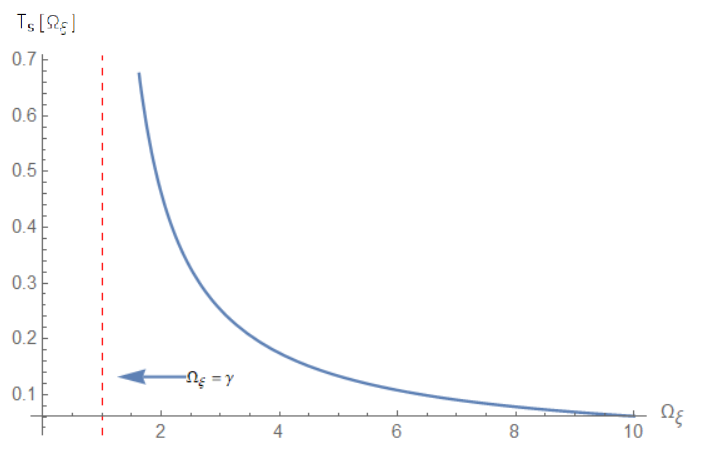}
\caption{Behavior of the time for singularities given by Eq. (\ref{TiempoBR}) as a function of $\Omega_{\xi}$, for the particular values of $\gamma=1$ and $\Omega_{\Lambda}=0.69$.}
\label{timeasfunctionomegachi}
\end{figure}

From Eq. (\ref{Tm1}) a singularity time, $T_{s}$, appears if we take $E\rightarrow\infty$, which gives
\begin{equation}\label{TiempoBR}
T_{s}=\frac{2\Omega_{\xi}\log \left[ \left(\frac{1-\sqrt{\Omega_{\Lambda}}}{1+\sqrt{\Omega_{\Lambda}}}\right)^{\frac{\gamma}{2\Omega_{\xi}\sqrt{\Omega_{\Lambda}}}}\left(1-\Omega_{\Lambda}\right)^{\frac{1}{2}}\left(\frac{-\Omega_{\xi}}{\gamma-\Omega_{\xi}}\right)\right]}{3\left(\gamma^{2}-\Omega^{2}_{\xi}\Omega_{\Lambda}\right)}.   
\end{equation}
At this future singularity, from Eqs. \eqref{tt}, \eqref{Pi} and the EoS equation, we can see that $\rho$, $p$ and $\Pi$ are divergent. If 
\begin{equation}\label{condicionBigRip}
    \Omega_{\xi}>\gamma,
\end{equation}
then the argument of the logarithm in Eq. (\ref{TiempoBR}) is always positive. Even more, if $\Omega_{\xi}={\gamma}/{\sqrt{\Omega_{\Lambda}}}>\gamma$, the numerator and denominator of the Eq. \eqref{TiempoBR} are zero, however
\begin{equation}
    \lim_{\Omega_{\xi}\rightarrow \frac{\gamma}{\sqrt{\Omega_{\Lambda}}}}{T_{s}}=\frac{\left(\sqrt{\Omega_{\Lambda} }-1\right) \log \left(\frac{1-\sqrt{\Omega_{\Lambda} }}{\sqrt{\Omega_{\Lambda} }+1}\right)-2 \sqrt{\Omega_{\Lambda} }}{6 \gamma  \left(\Omega_{\Lambda} -\sqrt{\Omega_{\Lambda} }\right)},
\end{equation}
i. e. $T_{s}$ is continue for $\Omega_{\xi}>\gamma$ and there are a change of sign in $\Omega_{\xi}=\gamma/\sqrt{\Omega_{\Lambda}}$ for both, the numerator and denominator in Eq. (\ref{TiempoBR}), yielding that $T_{s}$ is always positive because when $\Omega_{\xi}>{\gamma}/{\sqrt{\Omega_{\Lambda}}}$ the argument of the logarithm is lower than 1 (negative numerator) and the denominator is negative, as can be seen in Fig. \ref{timeasfunctionomegachi}, where $T_{s}$ given by Eq. \eqref{TiempoBR} is plotted as a function of $\Omega_{\xi}$. In Fig. \ref{graficoEdeT} the red dashed lines represents two times of singularities according to Eq. \eqref{TiempoBR}, for $\Omega_{\xi}=1.5$ and $\Omega_{\xi}=1.1$, where the time of singularities are $T=0.812997$, which is roughly equivalent to $11.6969$ Gyrs (0.86 times the lifetime of $\Lambda$CDM universe); and $T=2.26375$, corresponding to $32.5695$ Gyrs (2.4 times the lifetime of $\Lambda$CDM universe), respectively.

For $\Omega_{\xi}<\gamma$, there are no future singularities (no finite time is obtained from Eq. \eqref{TiempoBR}). From Eq. (\ref{Tm1}) we can see that $E(T)$ follows very close the behavior of standard model, ending with a de Sitter behavior at  $T\rightarrow +\infty$, which can be seen taking $E=\sqrt{\Omega_{\Lambda}}$ (equivalent to the solution  given by Eq. (\ref{HdeSitter})) in Eq. (\ref{Tm1}).

\begin{figure}[H]
\centering
\includegraphics[width=0.5\textwidth]{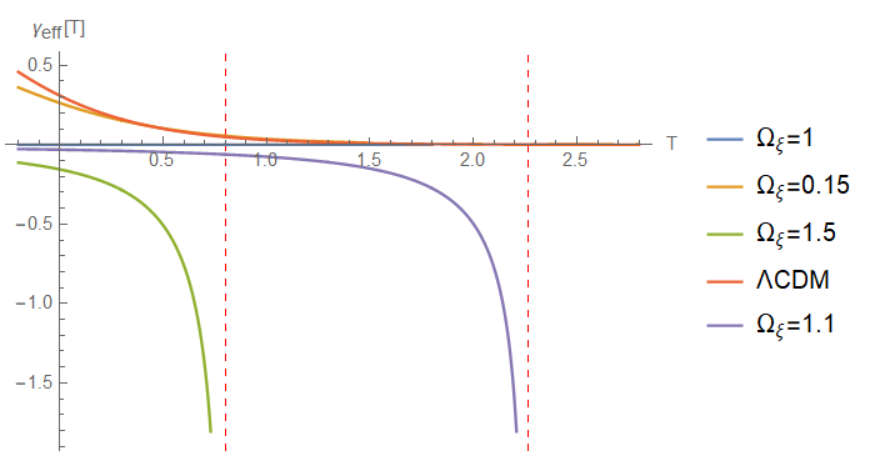}
\caption{Behavior of $\gamma_{eff}$ given by Eq. \eqref{weff} as a function of $T$ for the solutions with $m=1$, $\gamma=1$ and $\Omega_{\Lambda}=0.69$, for different values of $\Omega_{\xi}$. We also plotted $\gamma_{eff}$ for the $\Lambda$CDM model. The red dashed lines represent the times of singularities given by Eq. \eqref{TiempoBR} for $\Omega_{\xi}=1.5$ and $\Omega_{\xi}=1.1$, respectively.}
\label{graficowefectivo}
\end{figure}

In order to classify these singularities we need to explore the effective EoS of the models found. 
From Eqs. (\ref{Peff}), (\ref{ConsEq}) and the EoS equation one obtains as \cite{paperprofeAccelerated} that
\begin{equation}\label{we}
    \gamma_{eff}=\gamma+\frac{\Pi}{3H^{2}},
\end{equation}
and from Eq. \eqref{rr} it is possible to find an expression for the viscous pressure given by
\begin{equation}
    \Pi=-2\dot{H}-3\gamma H^{2},
\end{equation}
where using the above expression, we will have for (\ref{we}) the expression
\begin{equation}\label{weff0}
    \gamma_{eff}=-\frac{2\dot{H}}{3H^{2}},
\end{equation}
and using Eq. \eqref{Hpunto} in our dimensionless notation we will have
\begin{equation}\label{weff}
    \gamma_{eff}=\gamma-\Omega_{\xi}E+\frac{\Omega_{\xi}\Omega_{\Lambda}}{E}-\frac{\Omega_{\Lambda}\gamma}{E^{2}}.
\end{equation}
This $\gamma_ {eff}$ represents the effective EoS of a universe with a DE component modeled by a CC and a dissipative component. The phantom behavior of our solutions can be associated to the global composition of the universe. Fig. \ref{graficowefectivo} shows the behavior of $\gamma_{eff}$ as a function of $T$ for the solutions found, for different values of $\Omega_{\xi}$.

Let see now the type of singularities that we found in the dissipative CDM case ($\gamma=1$). For the solutions without singularities, i.e.,  $\Omega_{\xi}<1$, $\gamma_ {eff}$ evolves to $0$, representing the dominance of the CC at very far future times. In the solution with $\Omega_{\xi}>1$, $\rho$ and $p$ diverges and therefore, from Eq. \eqref{tt} $H$ and $a$ diverges, i. e., these solutions present Big-Rip singularities because $\gamma_{eff}$ from Eq. \eqref{weff} is always phantom, as can be seen from Fig. \ref{graficowefectivo}. It is important to note that since $H$ and $\dot{H}$ go to infinity for this singularity, then the Ricci scalar given by Eq. \eqref{ricci} also diverge. 

\subsection{Late-times singularities with $\Lambda<0$}

In this subsection we will study the singularities that arise from the solutions of Eq. \eqref{Hpunto} for a negative CC when $m=0$ and $m=1$. In this case, in order to compare with $\Lambda$CDM model, we will set in the general solutions $\gamma=1$ (CDM) and $\Omega_{\Lambda}=-0.69$. It is important to note from Eq. \eqref{rr} that the model with a negative CC can still gives an accelerated solution because of the negative pressure due to the bulk viscosity. Therefore, the election of $\Omega_{\Lambda}=-0.69$ is the first natural election in order to a further comparison, because, from Eqs. \eqref{tt} and \eqref{ConsEq} the usual Friedmann's constraint $\Omega_{m}+\Omega_{\Lambda}=1$ is not already valid and the values of $\Omega_{\Lambda}$ can, in principle, take any negative value.

\subsubsection{Cases for $m=0$}\label{solucionm0}
In this case we have three different types of solutions depending if the  discriminant, $\Delta_{0}$, given by Eq. (\ref{discriminant0}) is greater, equal, or lower than zero.

\textbf{(i) Case $\Delta_{0}>0$.} In this case the constraint for the values of a negative CC is  
\begin{eqnarray}\label{omegadeltacero1}
-\left(\frac{\Omega_{\xi}}{2\gamma}\right)^2<\Omega_{\Lambda}<0. 
\end{eqnarray}
We already have explained that this solution doesn't present any kind of singularity due to its bouncing behavior and the solution was already found in Eq. (\ref{H1}) (for the Hubble parameter) and in Eq. (\ref{factordescala1mcero}) (for the scale factor). It is interesting to mention that despite having a negative CC, this solution does not present Big-Crunch singularity, and at late times displays a de Sitter like expansion.

\textbf{(ii) Case $\Delta_{0}=0$.}  In this case the CC takes the particular value
\begin{eqnarray}\label{omegadeltacero}
\Omega_{\Lambda}=-\left(\frac{\Omega_{\xi}}{2\gamma}\right)^2,
\end{eqnarray}
and the solution for $E(T)$ takes the form
\begin{equation}\label{HtsolutionceroE}
E(T)=\frac{4+3\Omega_{\xi}(1-\frac{\Omega_{\xi}}{2\gamma})T}{4+6\gamma(1-\frac{\Omega_{\xi}}{2\gamma})T}.
\end{equation}
The corresponding  scale factor is given by 
\begin{equation}\label{a22}
    a(t)=\exp{\left[\frac{\Omega_{\xi}}{2\gamma}T\right]}\left[\frac{3\gamma}{2}T(1-\frac{\Omega_{\xi}}{2\gamma})+1\right]^{\frac{2}{3\gamma}}.
\end{equation}

It is straightforward to see from Eq. (\ref{HtsolutionceroE}) that if  $\Omega_{\xi}=2\gamma$, then $E=1$ for all time, corresponding to our de Sitter like solution given by \eqref{Hsolutiondeltacero} . If $\Omega_{\xi}>2\gamma$, $E$ goes to zero in a future time $T_{c}$, given by 
\begin{equation}\label{tiempoEcero}
    T_{c}=-\frac{4}{3\Omega_{\xi}\left(1-\frac{\Omega_{\xi}}{2\gamma}\right)}>0,
\end{equation}
which indicates that the scale factor takes a maximum value at this time and, from Eq. \eqref{a22}, goes to zero at a time given by 
\begin{equation}\label{tiempoinfinito2}
    T_{s}=-\frac{2}{3\gamma\left(1-\frac{\Omega_{\xi}}{2\gamma}\right)}>0.
\end{equation}
From Eq. \eqref{HtsolutionceroE} we can see that at the above time $E \rightarrow -\infty$, which means, from Eq. \eqref{tt}, that $\rho$ diverge and, from the EoS equation, $p$ diverge, indicating that in this case the future singularity corresponds to a Big-Crunch (Type OB singularity). It is important to note that since $H$ and $\dot{H}$ go to minus infinity for this singularity, then the Ricci scalar given by Eq. \eqref{ricci} also diverge. On the other hand, if $\Omega_{\xi}<2\gamma$, then $E>0$ for all time and goes to the value  $\Omega_{\xi}/2\gamma$ when $T \rightarrow+\infty$. Therefore, this solution asymptotically takes a de Sitter like behavior given by Eq. \eqref{Hsolutiondeltacero}. Note that for $\Omega_{\xi}\leq 2\gamma$ effectively we can drive the acceleration expansion of the universe when a negative CC is considered in our model, due only to the negativeness of the viscous pressure.
\begin{figure}[H]
\centering
\includegraphics[width=0.5\textwidth]{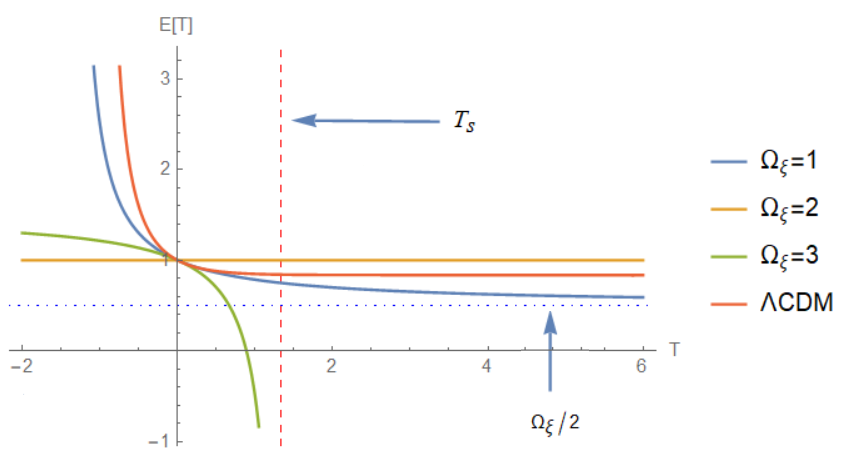}
\caption{Behavior of  $E(T)$,  given by Eq. \eqref{HtsolutionceroE} for different values of $\Omega_{\xi}$ and for the particular value of $\gamma=1$. $\Omega_{\Lambda}$ is given by Eq. \eqref{omegadeltacero}. We also plotted the $\Lambda$CDM model. The red dashed line represent the  singularity time given by Eq. \eqref{tiempoinfinito2} for $\Omega_{\xi}=3$.}
\label{graficomcerodeltacero}
\end{figure}

In Fig. \ref{graficomcerodeltacero} we display the behavior of the Hubble parameter \eqref{HtsolutionceroE} for $\gamma=1$. The Big-Crunch singularity appears for the particular values of $\gamma=1$ and $\Omega_{\xi}=3$ evaluated in Eq. \eqref{tiempoinfinito2}, and leads to $T_{s}=4/3$, which is roughly equivalent to $19.18$ Gyrs (1.45 times the lifetime of the $\Lambda$CDM universe).

\begin{figure}[H]
\centering
\includegraphics[width=0.5\textwidth]{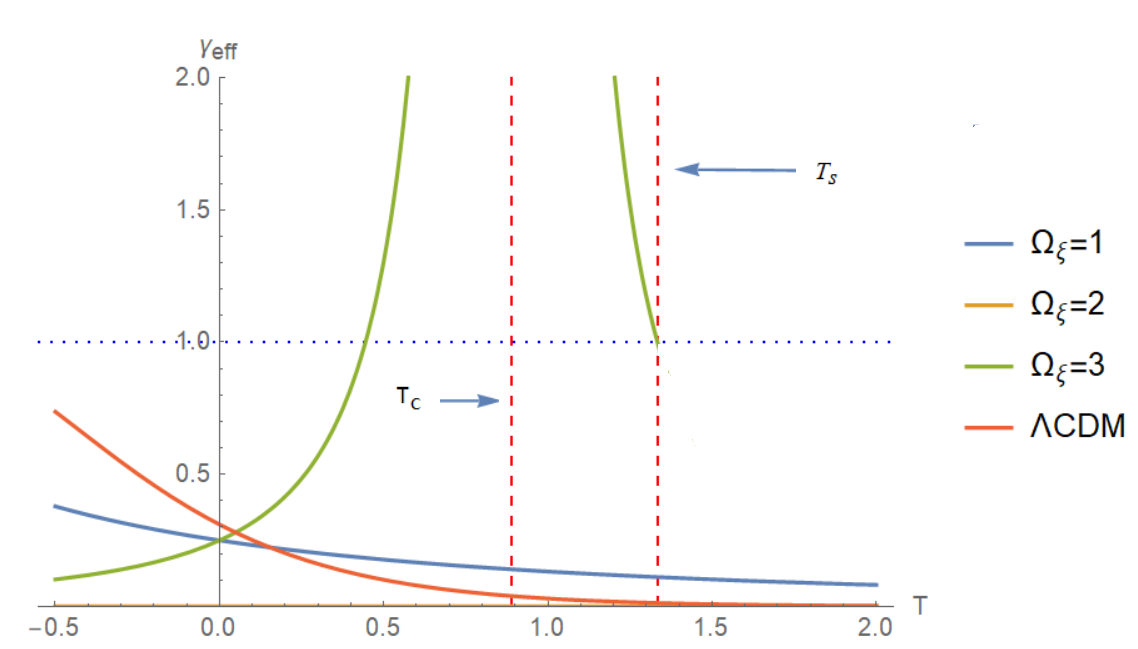}
\caption{Behavior of $\gamma_{eff}$  given by Eq.  \eqref{gammaeffmcero1} as a function of $T$ for the solution with $m=0$ and $\Delta_{0}=0$, for the particular value of $\gamma=1$ and for $\Omega_{\Lambda}$ given by Eq. \eqref{omegadeltacero}, for different values of $\Omega_{\xi}$. $T_{c}$ and $T_{s}$ are given by \eqref{tiempoEcero} and Eq. \eqref{tiempoinfinito2}, respectively. We also plotted the $\gamma_{eff}$ for the $\Lambda$CDM model.}
\label{graficomcerogamma}
\end{figure}

From Eq. \eqref{weff0} the effective barotropic index for this solution is
\begin{equation}\label{gammaeffcero}
    \gamma_{eff}=\gamma-\frac{\Omega_{\xi}}{E}+\frac{\left|\Omega_{\Lambda}\right|\gamma}{E^{2}},
\end{equation}
and from the solution given by Eq. \eqref{HtsolutionceroE}, we have
\begin{equation}\label{gammaeffmcero1}
    \gamma_{eff}=\frac{16\gamma\left(\frac{\Omega_{\xi}}{2\gamma}-1\right)^{2}}{\left(4-3\Omega_{\xi}T\left(\frac{\Omega_{\xi}}{2\gamma}-1\right)\right)^{2}}.
\end{equation}
Note that, if we substitute Eq. \eqref{tiempoEcero} (where $E$=0 and $a$ takes his maximum value) in Eq. \eqref{gammaeffmcero1}, we will get $\gamma_{eff}\rightarrow+\infty$, and if we substitute Eq. \eqref{tiempoinfinito2} (Big Crunch time) we will get $\gamma_{eff}=\gamma$ (according to Eq. \ref{gammaeffcero}). The behavior of this $\gamma_{eff}$ is presented in the Fig. (\ref{graficomcerogamma}).

It is important to mention that as the viscosity increases, the value of $\Omega_{\Lambda}$ also increases, which can be seen from Eq. \eqref{omegadeltacero}; also the time $T_{c}$, where the scale factor takes its maximum value, occurs after the current time.

\textbf{(iii) Case $\Delta_{0}<0$}. In this case the dimensionless density parameter associated to the negative CC satisfied the following inequality 
\begin{equation}\label{lambdanegativo2}
 \Omega_{\Lambda}<-\left(\frac{\Omega_{\xi}}{2\gamma}\right)^2,
\end{equation}
and the exact solution takes the following form
\begin{eqnarray}\label{H3}
 E(T)&=&-\frac{\sqrt{\left | \bar{\Delta}_{0}\right|}}{2}\tan\left(\frac{3\gamma \sqrt{\left | \bar{\Delta}_{0}\right|}T}{4}-\arctan\left(\frac{2-\frac{\Omega_{\xi}}{\gamma}}{\sqrt{\left | \bar{\Delta}_{0}\right|}}\right)\right)\nonumber \\
&+&\frac{\Omega_{\xi}}{2\gamma},
\end{eqnarray}
with a scale factor given by
\begin{equation}\label{a03}
    a(T)=\exp{\left(\frac{\Omega_{\xi}}{2\gamma}T\right)}\left\{\frac{\cosh{\left[\frac{3\gamma\sqrt{\left|\bar{\Delta}_{0}\right|}}{4}T-\arctan{\left(\frac{2-\frac{\Omega_{\xi}}{\gamma}}{\sqrt{\left|{\bar{\Delta}}_{0}\right|}}\right)} \right]}}{\cos\left[\arctan{\left(\frac{2-\frac{\Omega_{\xi}}{\gamma}}{\sqrt{\left|\bar{\Delta}_{0}\right|}}\right)}\right]}\right\}^{\frac{2}{3\gamma}},
\end{equation} 

\begin{figure}[H]
\centering
\includegraphics[width=0.5\textwidth]{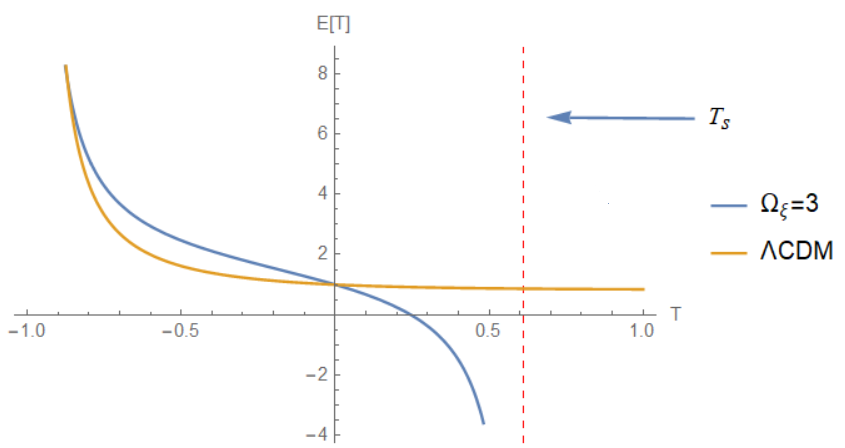}
\caption{Behavior of  $E(T)$ given by Eq. \eqref{H3}, for different values of $\Omega_{\xi}$ and for the particular values $\gamma=1$, $\Omega_{\Lambda}=-4$ and  $\Omega_{\xi}=3$, according to restriction \eqref{lambdanegativo2}. We also plotted the  $\Lambda$CDM model. The red dashed line represent the  singularity time given by Eq. \eqref{expresion2tiempo}.}
\label{Edeltamenoracero}
\end{figure}

In order to explore the possibility of future singularities, we found from Eq. \eqref{H3} $T$ as a function of $E$, obtaining
\begin{eqnarray}\label{Tdeltamenoracero}
&&T(E)=\frac{4}{3\gamma\sqrt{\abs{\bar{\Delta}_{0}}}}\times \nonumber \\
&&\left(\arctan\left({\frac{\frac{\Omega_{\xi}}{\gamma}-E}{\sqrt{\left|\bar{\Delta}_{0}\right|}}}\right)+\arctan\left(\frac{2-\frac{\Omega_{\xi}}{\gamma}}{\sqrt{\left|\bar{\Delta}_{0}\right|}}\right)\right).
\end{eqnarray}
From this equation we can notice that $E$ is zero in a time given by
\begin{eqnarray}\label{tiempoBCdeltamenoracero}
&&T_{c}=\frac{4}{3\gamma\sqrt{\abs{\bar{\Delta}_{0}}}}\times \nonumber \\
&&\left(\arctan\left({\frac{\frac{\Omega_{\xi}}{\gamma}}{\sqrt{\left|\bar{\Delta}_{0}\right|}}}\right)+\arctan\left(\frac{2-\frac{\Omega_{\xi}}{\gamma}}{\sqrt{\left|\bar{\Delta}_{0}\right|}}\right)\right),
\end{eqnarray}
which indicates that the scale factor takes a maximum value at this time and goes to zero when $E\rightarrow-\infty$, as can be seen from Eq. \eqref{Tdeltamenoracero}, in a time given by
\begin{eqnarray}\label{expresion2tiempo}
T_{s}=\frac{4}{3\gamma\sqrt{\abs{\bar{\Delta}_{0}}}}\times
\left(\frac{\pi}{2}+\arctan\left(\frac{2-\frac{\Omega_{\xi}}{\gamma}}{\sqrt{\left|\bar{\Delta}_{0}\right|}}\right)\right),
\end{eqnarray}
i. e., $a\rightarrow 0$, as can be shown if we substitute the time given in Eq. \eqref{expresion2tiempo} in \eqref{a03}. So, from Eq. \eqref{tt} $\rho$ diverge and from the EoS equation $p$ also diverge. Therefore, at this time occurs a Big-Crunch singularity (Type 0B). It is important to note that since $H$ and $\dot{H}$ go to minus infinity for this singularity, then the Ricci scalar given by Eq. \eqref{ricci} also diverge.
This is the only scenario that we have for this solution and a de Sitter asymptotic expansion is not possible, as it can be check from Eq. \eqref{Hsolutionpositive}. In Fig. \ref{Edeltamenoracero} we present the behavior for $E$ as a function of $T$, given by Eq.\eqref{H3}. The value of time of singularity shown in this figure is  $T= 0.609495$, which is roughly equivalent to $8.76906$Gyrs (0.64 times the life of the $\Lambda$CDM universe).

\begin{figure}[H]
\centering
\includegraphics[width=0.5\textwidth]{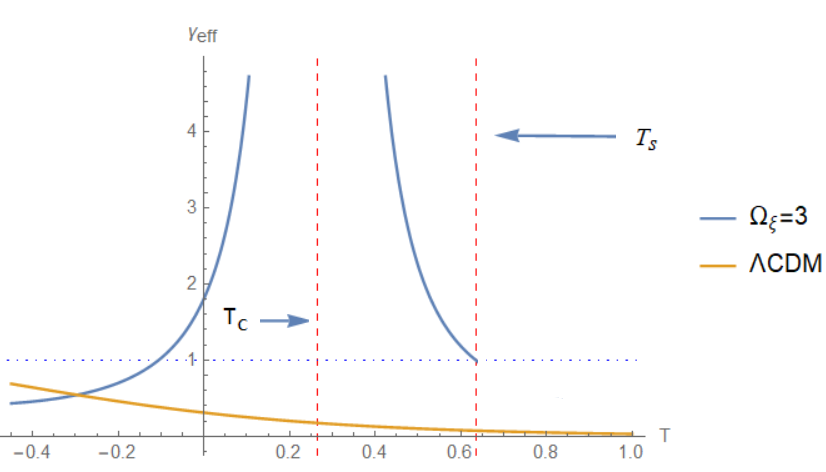}
\caption{Behavior of $\gamma_{eff}(T)$,  given by Eq. \eqref{H3}, for the solution with $m=0$ and $\Delta_{0}<0$, for the particular values of $\gamma=1$ and $\Omega_{\Lambda}=-4$. We use $\Omega_{\xi}=3$ according to restriction \eqref{lambdanegativo2}. $T_{c}$ and $T_{s}$ are given by \eqref{tiempoBCdeltamenoracero} and Eq. \eqref{expresion2tiempo}, respectively. We also plotted $\gamma_{eff}$ for the $\Lambda$CDM model.}
\label{gammadeltamenoracero}
\end{figure}

For the solution given by Eq. \eqref{H3} we have, from Eq. \eqref{gammaeffcero}, that
\begin{flalign}\label{gammadeltanegativo}
\gamma_{eff}= 
 \frac{\gamma  \left|\bar{\Delta} _0\right| \sec\left(\frac{3}{4} \gamma  \sqrt{\left|\bar{\Delta} _0\right|} T-\arctan\left(\frac{2-\frac{\Omega _{\xi }}{\gamma }}{\sqrt{\left|\bar{\Delta} _0\right|}}\right)\right)^{2}}{4 \left(\frac{\Omega _{\xi }}{2 \gamma }-\frac{\sqrt{\left|\bar{\Delta} _0\right|}}{2}  \tan \left(\frac{3}{4} \gamma  \sqrt{\left|\bar{\Delta} _0\right|} T-\arctan\left(\frac{2-\frac{\Omega _{\xi }}{\gamma }}{\sqrt{\Delta _0}}\right)\right)\right)^{2}},
\end{flalign}
Note that, if we substitute Eq. \eqref{tiempoBCdeltamenoracero} (where $E$=0) in Eq. \eqref{gammadeltanegativo} we will get $\gamma_{eff}\rightarrow+\infty$, and if we substitute Eq. \eqref{expresion2tiempo} we will get $\gamma_{eff}=\gamma$ (according to Eq. \ref{gammaeffcero}).  The behavior of the above expression is presented in Fig. \ref{gammadeltamenoracero}.

\subsubsection{Cases for $m=1$}\label{casom1negativo}

\begin{figure}[H]
\centering
\includegraphics[width=0.5\textwidth]{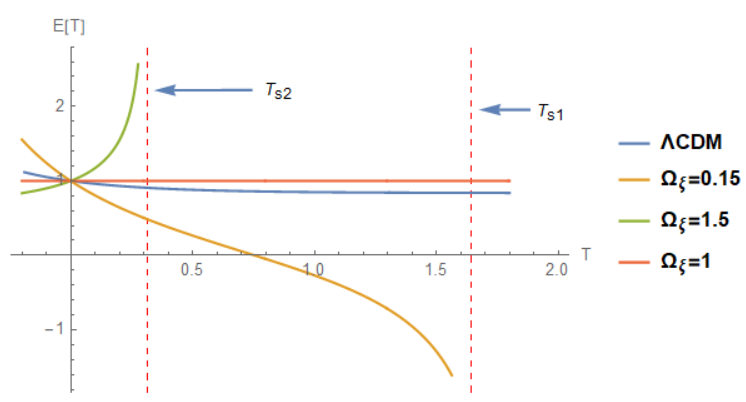}
\caption{{Numerical} behavior of $E(T)$, given by Eq. \eqref{Tm2}, for different values of $\Omega_{\xi}$ and for the particular values of $\gamma=1$ and $\Omega_{\Lambda}=-0.69$. We also plotted the $\Lambda$CDM model. The red dashed line represent $T_{s1}$ and $T_{s2}$  given by Eq. \eqref{m1BC} and Eq. \eqref{TiempoBR2} respectively.}
\label{Figuram1negativo}
\end{figure}

In this case the solution is given by

\begin{eqnarray}\label{Tm2}
&T(E)=\frac{2\Omega_{\xi}  \sqrt{\abs{\Omega_{\Lambda}}} \log \left(\frac{(1+\abs{\Omega_{\Lambda}} )^{\frac{1}{2}} (\gamma - E\Omega_{\xi} )}{\left(E^2+\abs{\Omega_{\Lambda}} \right)^{\frac{1}{2}} (\gamma -\Omega_{\xi} )}\right)}{{3 \sqrt{\abs{\Omega_{\Lambda}} } \left(\gamma^2+\Omega^{2}_{\xi} \abs{\Omega_{\Lambda}} \right)}} \nonumber \\
& +\frac{2\gamma\left(\arctan{\left(\frac{1}{\sqrt{\abs{\Omega_{\Lambda}}}}\right)}-\arctan{\left(\frac{E}{\sqrt{\abs{\Omega_{\Lambda}}}}\right)}\right)}{{3 \sqrt{\abs{\Omega_{\Lambda}} } \left(\gamma^2+\Omega^{2}_{\xi} \abs{\Omega_{\Lambda}} \right)}}.
\end{eqnarray}
Note that $T\rightarrow \infty, \;\forall E$ when $\Omega_{\xi}=\gamma$, in other words, this case represents the de Sitter case  given by Eq. \eqref{H1deSitter}, that is $H(t)=H_{0}$, $\forall t$, as it can be seen from Fig. \ref{Figuram1negativo}. If $\Omega_{\xi}<\gamma$, $E=1$ at $T=0$ and goes to zero in a future time $T_{c}$ given by
\begin{eqnarray}\label{TiempoBigcrunch}
&T_{c}=\frac{2\Omega_{\xi}  \sqrt{\abs{\Omega_{\Lambda}}} \log \left(\frac{(1+\abs{\Omega_{\Lambda}} )^{\frac{1}{2}} (\gamma )}{\left({\sqrt{\abs{\Omega_{\Lambda}}}} \right) (\gamma -\Omega_{\xi} )}\right)}{{3 \sqrt{{\abs{\Omega_{\Lambda}}} } \left(\gamma^2+\Omega^{2}_{\xi} \abs{\Omega_{\Lambda}} \right)}} \nonumber \\
& +\frac{2\gamma\left(\arctan{\left(\frac{1}{\sqrt{\abs{\Omega_{\Lambda}}}}\right)}\right)}{{3 \sqrt{\abs{\Omega_{\Lambda}} } \left(\gamma^2+\Omega^{2}_{\xi} \abs{\Omega_{\Lambda}} \right)}}, 
&\end{eqnarray}
indicating  that  the  scale  factor  takes  a  maximum value at this time, and goes to zero when $E\rightarrow-\infty$ at a time given by

\begin{eqnarray}\label{m1BC}
&T_{s1}=\frac{2\Omega_{\xi}\log \left(\left(1+\abs{\Omega_{\Lambda}}\right)^{\frac{1}{2}}\left(\frac{\Omega_{\xi}}{\gamma-\Omega_{\xi}}\right)\right)}{3\left(\gamma^{2}+\Omega^{2}_{\xi}\abs{\Omega_{\Lambda}}\right)}
\nonumber \\
& +\frac{2\gamma\left(\arctan{\left(\frac{1}{\sqrt{\abs{\Omega_{\Lambda}}}}\right)}+\frac{\pi}{2}\right)}{{3 \sqrt{\abs{\Omega_{\Lambda}} } \left(\gamma^2+\Omega^{2}_{\xi} \abs{\Omega_{\Lambda}} \right)}}.
&\end{eqnarray}
Therefore, if $a\rightarrow0$ and since $E\rightarrow-\infty$, from Eq. \eqref{tt} $\rho$ diverge and from the Eos $p$ also diverge, so this solution represents a universe with a Big-Crunch type future singularity (Type 0B) and $\gamma_{eff}$ from \eqref{weff} goes to infinity. It is important to note that since $H$ and $\dot{H}$ go to minus infinity for this singularity, then the Ricci scalar given by Eq. \eqref{ricci} also diverge. In  Fig. \ref{Figuram1negativo} \textbf{we have numerically found the behavior of $E$ as a function of $T$}, given by Eq. \eqref{Tm2}. The value of time of singularity show in this figure is $T_{s1}= 1.79003$, which is roughly equivalent to $25.7539$Gyrs (1.87 times the life of the $\Lambda$CDM universe).

\begin{figure}[H]
\centering
\includegraphics[width=0.5\textwidth]{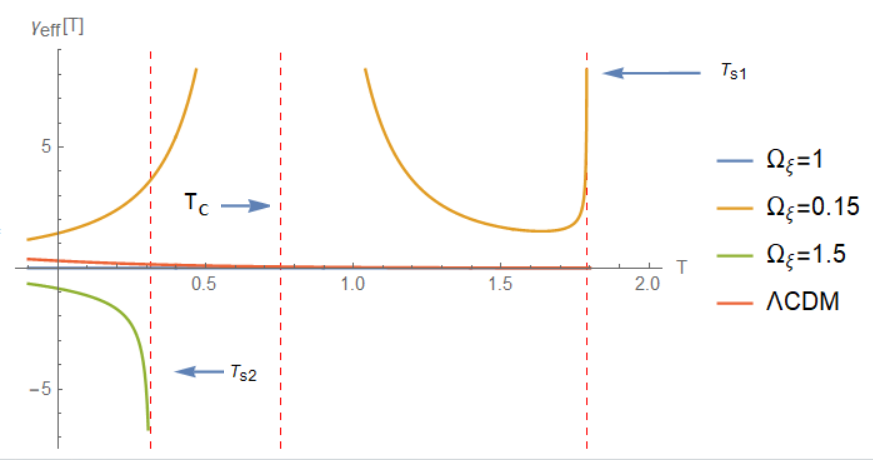}
\caption{Behavior of $\gamma_{eff}$  given by Eq. \eqref{weff} as a function of $T$ for the solutions with $m=1$, for the particular values of $\gamma=1$ and $\Omega_{\Lambda}=-0.69$, and different values of $\Omega_{\xi}$. We also plotted  $\gamma_{eff}$ for the $\Lambda$CDM model. $T_{c}$, $T_{s1}$ and $T_{s2}$ are given by Eq. \eqref{TiempoBigcrunch}, Eq. \eqref{m1BC} and Eq. \eqref{TiempoBR2} respectively.}
\label{graficowefectivonegativo}
\end{figure}

If $\Omega_{\xi}>\gamma$, $E>0$ for all time and goes to infinite in a finite time given by
\begin{eqnarray}\label{TiempoBR2}
&T_{s2}=\frac{2\Omega_{\xi}\log \left(\left(1+\abs{\Omega_{\Lambda}}\right)^{\frac{1}{2}}\left(\frac{-\Omega_{\xi}}{\gamma-\Omega_{\xi}}\right)\right)}{3\left(\gamma^{2}+\Omega^{2}_{\xi}\abs{\Omega_{\Lambda}}\right)}
\nonumber \\
& +\frac{2\gamma\left(\arctan{\left(\frac{1}{\sqrt{\abs{\Omega_{\Lambda}}}}\right)}-\frac{\pi}{2}\right)}{{3 \sqrt{\abs{\Omega_{\Lambda}} } \left(\gamma^2+\Omega^{2}_{\xi} \abs{\Omega_{\Lambda}} \right)}}. 
\end{eqnarray}
As in the case of a positive CC, Eq. \eqref{TiempoBR2} is always positive for any value of $\Omega_{\xi}>\gamma$. Now, if we substitute $\gamma=1$ (dust case) and if we use $\Omega_{\Lambda}=-0.69$ (to compare with the case of positive CC), then for $\Omega_{\xi}>1$, $\rho$ and $p$ diverges and therefore, from Eq. \eqref{tt} $H$ and $a$ diverges, i. e., these solutions present Big-Rip singularities because $\gamma_{eff}$ from \eqref{weff} is always phantom, as can be seen from Fig. \ref{graficowefectivonegativo}. In this figure for $\Omega_{\xi}=1.5$, we have a Big Rip singularity time of $T= 0.315246$, which is roughly equivalent to $4.53558$Gyrs (0.33 times the life of the $\Lambda$CDM universe). It is important to note that since $H$ and $\dot{H}$ go to infinity for this singularity, then the Ricci scalar given by Eq. \eqref{ricci} also diverge. 

\subsection{\textbf{Early-Time singularities for the case of $\Lambda>0$}}

In the case of early-singularities we explore the behavior of our exact solutions backward in time, taking $\gamma=4/3$ (radiation) or even $\gamma \leq 2$ (cuasi stiff fluid), assuming that some kind of dissipation is possible at these very early stages. As an initial condition for our solutions, we will assume that $\Omega_{radiation}$ takes values very close to one, which is reasonable to assume during the radiation dominant era. In order to make comparisons we will consider an early evolution stage of the $\Lambda$CDM model.  Our model is based on the composition of only two fluids, (i) dissipative matter (ii) dark energy modeled as a CC. According to our previous discussion, radiation is imposed as the dominant fluid in relation to the value of $\Omega_{\Lambda}$, therefore from Eq. (4) for an arbitrary very early radiation time we can consider the value of $\Omega_{\Lambda}=10^{-6}$, in order to use the exact solution found and explore its behavior to the past. On the contrary, during the current DE era the actual value of radiation density, according to observation, is $\Omega_{radiation}=9.72\times10^{-5}$ \cite{WMAP2013,planck2013}

The below discussion correspond to the case of a dissipative radiation fluid. The initial condition chosen, $T=H_{0}t=0$, represents the arbitrary moment during the radiation dominance when $\Omega_{\Lambda}=10^{-6}$ and $1-\Omega_{\Lambda}=\Omega_{radiation}$ is very close to one. Here $H_{0}$ and $a_{0}=1$ are the Hubble parameter and the scale factor at this arbitrary moment and we keep the definition for $E(T)$. Clarifying these new particular initial conditions, we can use the solutions previously found looking their behavior backward in time. The value of $\Omega_{\xi}$  represents then the dimensionless density of dissipation at this initial time chosen above. 

\subsubsection{Case for $m=0$}
We have found that the only solution for a positive CC is given, in this case, when the discriminant of Eq. \eqref{discriminant0} is positive, but this solution presents a bouncing behavior as it was discussed in section \ref{prueba}, so this solution describe a regular universe without an early singularity.

\subsubsection{Case for $m=1$}\label{earlym1}
The general solution for an arbitrary $\gamma$ for this case corresponds to the expression \eqref{Tm1}. The expression \eqref{TiempoBR} corresponds a time when the energy density and $E$ tends to infinity, which are the same conditions required to have an early Type 0A (Big-Bang) singularity, with the difference that, in this case, the scale factor tends to zero. We already have discussed analytically and graphically (see Fig. \ref{timeasfunctionomegachi}) Eq. \eqref{TiempoBR}, showing that is strictly positive, so this universe doesn't have early singularities.  We will later discuss in detail in section \ref{soft} the behavior of this solution at early times.

\subsection{Early time singularity for the case of $\Lambda<0$}

\subsubsection{Cases for $m=0$}
For this case the only solutions that present a singularities are those with a discriminant equal or less than zero; recall when $\Delta_{0}>0$ the solution has a Bouncing-type behavior given by \eqref{factordescala1mcero}.

\begin{figure}[H]
\centering
\includegraphics[width=0.5\textwidth]{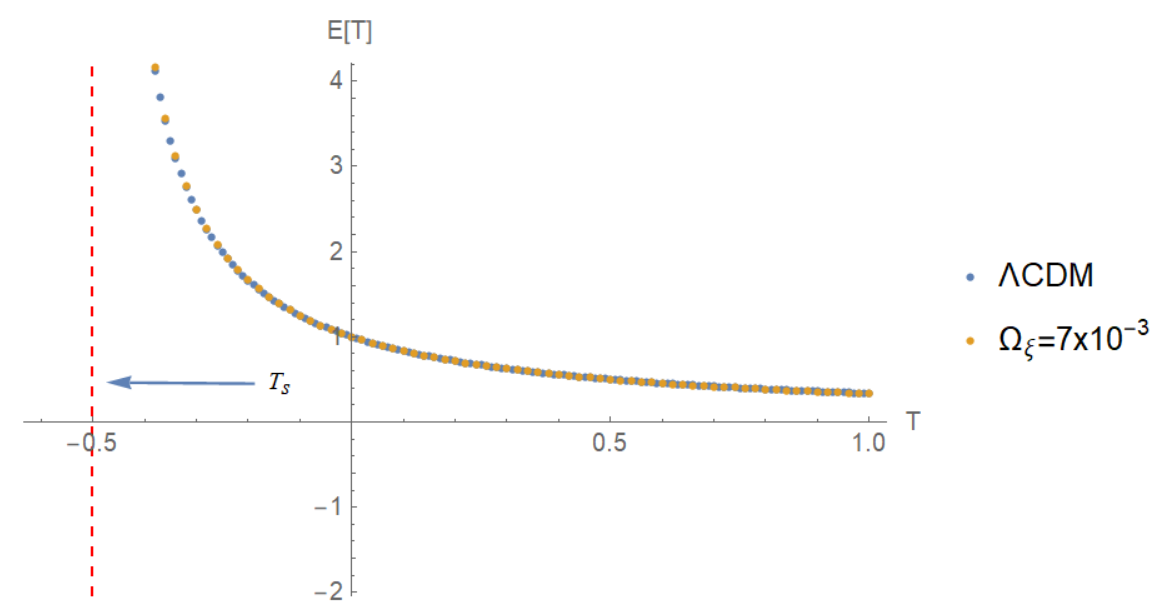}
\caption{ Behavior of  $E(T)$,  given by Eq. \eqref{HtsolutionceroE} for $\gamma=4/3$, $\Omega_{\Lambda}=-10^{6}$. $\Omega_{\xi}$ is restricted by Eq. \eqref{omegadeltacero}. We also plotted the $\Lambda$CDM model. The red dashed line represent the  singularity time given by Eq. \eqref{tiempoinfinito2}}
\label{graficomcerodeltaceronegativo}
\end{figure}

\textbf{(i) Case $\Delta_{0}=0$.} We consider the behavior backwards in time of expression $E(T)$ and $a(T)$ given by \eqref{HtsolutionceroE} and \eqref{a22}, respectively. Even more, for this solution the time for singularity is given by (\ref{tiempoinfinito2}), resulting in a scale factor of null value, and since $H$ and $\dot{H}$ go to infinity for this singularity, then the Ricci scalar given by Eq. \eqref{ricci} also diverge. To get a early singularity we have a restriction for $\Omega_{\xi}$ from \eqref{tiempoinfinito2} given by $\Omega_{\xi}<8/3$ for the case of radiation. In this sense, $\rho$ and $p$ diverges, so we will get a Type $0A$ singularity (Big-Bang). From the value of the CC, Eq. (\ref{omegadeltacero}) leads to $\Omega_{\xi}=7\times 10^{-3}$. In Fig. \ref{graficomcerodeltaceronegativo} we present the behavior for $E$ as a function of $T$, given by Eq.\eqref{H3}. 

\begin{figure}[H]
\centering
\includegraphics[width=0.5\textwidth]{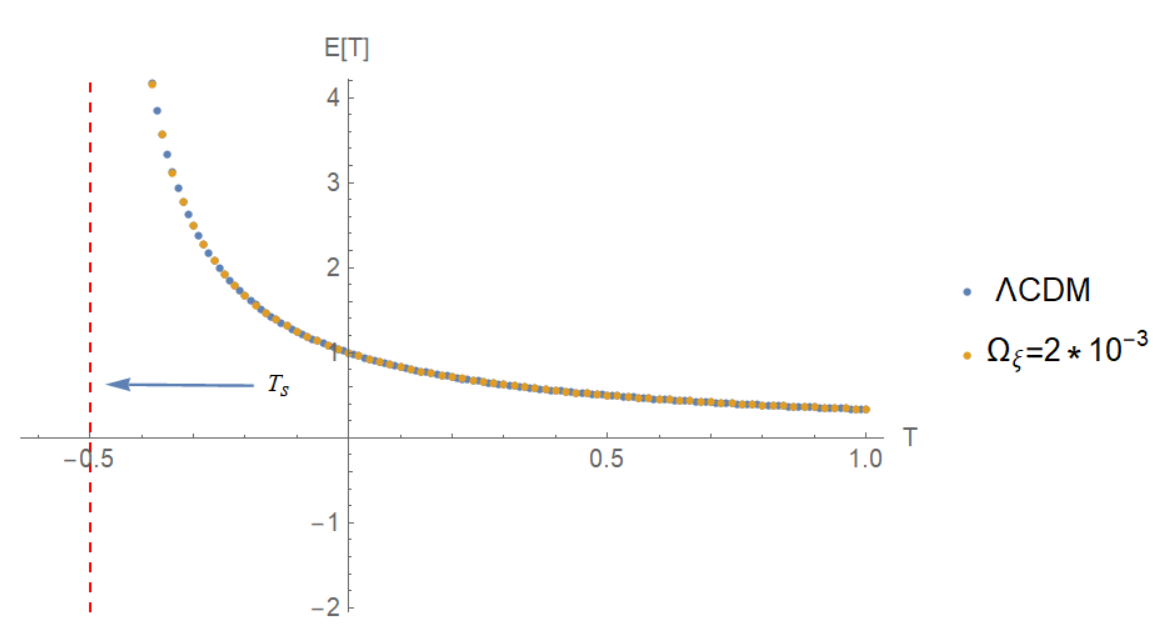}
\caption{Behavior of  $E(T)$ given by Eq. \eqref{H3},  for the particular values $\gamma=4/3$, $\Omega_{\Lambda}=-10^{-6}$ and  $\Omega_{\xi}=2\times 10^{-3}$, according to restriction \eqref{lambdanegativo2}. We also plotted the  $\Lambda$CDM model. The red dashed line represent the  singularity time given by Eq. \eqref{expresion3tiempo}.}
\label{Edeltamenoraceronegativo}
\end{figure}

\textbf{(ii) Case $\Delta_{0}<0$.} We consider the behavior backwards in time of expression $E(T)$ and $a(T)$ given by Eqs. \eqref{H3} and \eqref{a03}, respectively. Even more, for this solution the time for singularity is given by (\ref{expresion2tiempo}), resulting in a scale factor of null value, and since $H$ and $\dot{H}$ go to infinity for this singularity, then the Ricci scalar given by Eq. \eqref{ricci} also diverge. For early time singularity  we need to considered, from  \eqref{Tdeltamenoracero}, $E\rightarrow+\infty$  to get
\begin{eqnarray}\label{expresion3tiempo}
T_{s}=\frac{4}{3\gamma\sqrt{\abs{\bar{\Delta}_{0}}}}\times
\left(-\frac{\pi}{2}+\arctan\left(\frac{2-\frac{\Omega_{\xi}}{\gamma}}{\sqrt{\left|\bar{\Delta}_{0}\right|}}\right)\right).
\end{eqnarray}
From the previous expression $\rho$ and $p$ diverges, so this is a type 0A singularity (Big-Bang). The value of $\Omega_{\Lambda}=-10^{-6}$  leads to $\Omega_{\xi}<3\times10^{-3}$ from Eq. \eqref{lambdanegativo2}.
In Fig. \ref{Edeltamenoraceronegativo} we present the behavior for $E$ as a function of $T$, given by Eq.\eqref{H3}.

\subsubsection{case $m=1$}
We discuss in seccion \ref{casom1negativo} that the time for singularity is given by \eqref{TiempoBR2} and is strictly positive, so this solution as in the case of positive CC  doesn't have a singularity in early stages either. A detailed discussion about this behavior will be done in section \ref{soft}

\section{Soft-Big Bang}\label{soft}

\begin{figure}[H]
\centering
\includegraphics[width=0.5\textwidth]{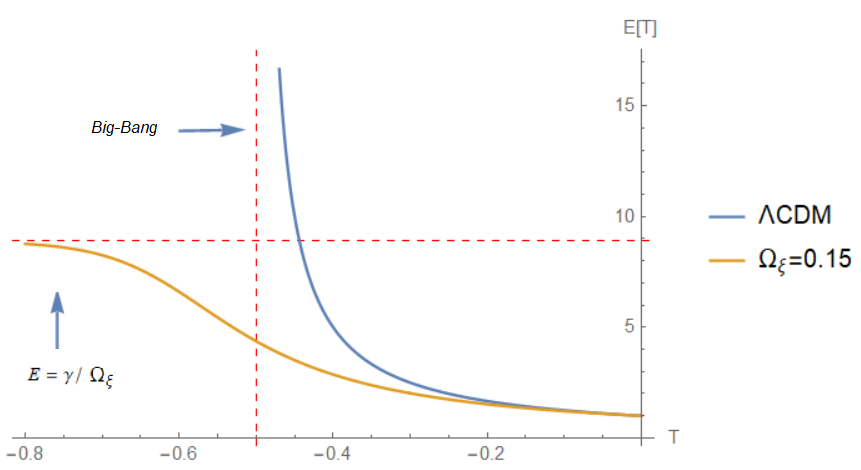}
\caption{{Numerical} behavior for $E(T)$ given by Eq. \eqref{Tm1} for $m=1$, $\Omega_{\Lambda}=10^{-6}$ and $\gamma=4/3$. We also plotted the behavior of the $\Lambda$CDM model.}
\label{figurasinsingularidad}
\end{figure}

As we have discussed in section \ref{earlym1}, the solution given by Eq. (\ref{Tm1}) (case with $m=1$ and $\Omega_{\Lambda}>0$), when $\Omega_{\xi}<\gamma$,  describe a universe without initial singularity. In this particular solution, when $T\rightarrow-\infty$, we obtain that the Hubble parameter is given by Eq. (\ref{H1deSitter}), and the same behavior is obtained if we considered a negative CC as can be seen in Eq.\eqref{Tm2} . In Fig. \ref{figurasinsingularidad} we have numerically found the behavior of $E$ as a function of $T$.

\begin{figure}[H]
\centering
\includegraphics[width=0.5\textwidth]{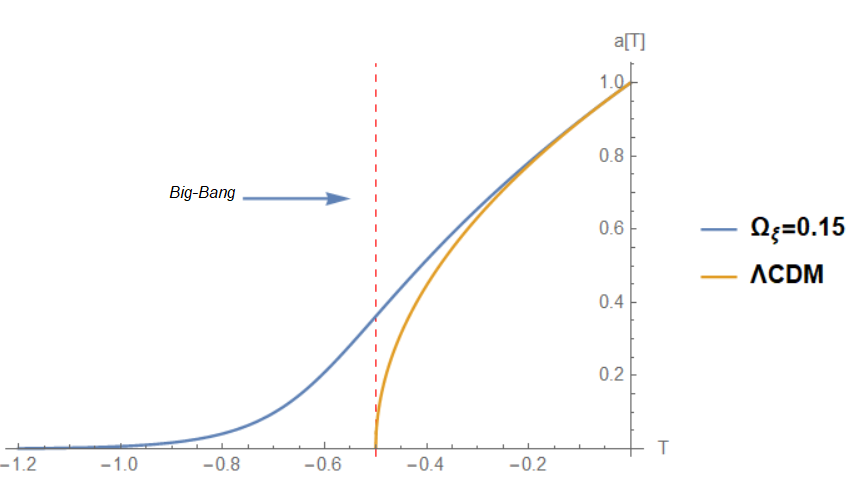}
\caption{Behavior of numerical integration from Eq. \eqref{Tm1} to get $a(T)$ for $m=1$, $\Omega_{\Lambda}=10^{-6}$ and $\gamma=4/3$. We also plotted the behavior of the $\Lambda$CDM model.}
\label{asoft}
\end{figure}

Note that taking the limit $\Omega_{\xi}\rightarrow 0$ in Eq. (\ref{H1deSitter}) we obtain that $E\rightarrow \infty$, which is the behavior corresponding to a Big Bang singularity in the past. Hence, this solution turns into a $\Lambda$CDM model with a Big-Bang singularity when dissipation is neglected. The behavior of the scale factor is shown in Fig. \ref{asoft}.

At $T\rightarrow-\infty$, $H\rightarrow \gamma/(3\xi_{0})$ and $\dot{H}\rightarrow 0$, so the Ricci scalar given by Eq. (\ref{ricci}) takes the value
\begin{equation}
    R=\frac{4\gamma^{2}}{3\xi^{2}_{0}},
\end{equation}
indicating that there is no curvature singularity in this solution. In the infinity past $a=0$ and $H$ takes a constant value. If $\xi_{0}\rightarrow 0$ the behavior of the standard model is recovered with $R\rightarrow \infty$ when $a=0$ in some finite time in the past. 
The inclusion of dissipation without a CC led to these soft-Big Bang scenarios \cite{Big.Bang}.

This solution is different from the soft-Big Bang studied in \cite{Softbang1,SoftBang2} or from other singularity-free models these suggested by Israel $\&$ Rosen \cite{Rosen}, or by Blome $\&$ Priester \cite{BigBounce} where the universe begin from either by a tiny bubble in a homogeneous and isotropic quantum state with the diameter of a Planck length as an initial condition, or start from an Einstein static universe, with a radius determined by the value of $\Lambda$, before entering a never-ending period of de Sitter expansion. The solution discussed in \cite{Softbang1} has the particularity of having a finite scale factor in the infinite past.

\section{Conclusions and final discussions}

We have discussed throughout this work the late and early times behavior of the exact solutions of viscous $\Lambda$CDM models, looking for the conditions to have future and past singularities, following the classification given in \cite{tiposdeBigRip,clasificacionBigRipdetallada}, and we have also found the possibility of solutions describing regular universes. In the late time model we consider a universe filled with dissipative CDM and CC and in the early time model we consider a universe filled with dissipative radiation and CC, taking into consideration two different expression for the dissipation, a constant bulk viscosity and a bulk viscosity proportional to the energy density. We extend this study for a dissipative fluid model with a negative CC. In the Table \ref{tabla1} we summarize the early and late time singalirities obtained in each solutions and in Table \ref{tabla2} we summarize the asymptotic early and late behavior without singularities found for these models.


For a positive CC in the late time behavior a remarkable result of the solution with $m=1$ and $\Omega_{\xi}<1$ is that the solution behaves at late times like the de Sitter model, regardless of the viscosity value. In this sense this solution is suitable to constraint with the cosmological data, knowing that it evolves very close to the $\Lambda$CDM model. 


For a negative CC in the late time behavior a remarkable result for the solution with $m=1$ is that the dissipation in the DM component can drive the accelerated expansion and even a future Big-rip singularity, avoiding the big crunch singularity, that occurs for a flat DM filed universe with negative CC.

It is important to mention that within the context of singularities in phantom DE, the little rip singularity are discussed in the literature \cite{littleRIp,littlerip2,brevik2} under the context of having a universe in which the DE density increases without bound and the universe never reaches a finite-time for singularity. In our work these type of singularities don't appear because we are not considering phantom DE and unlike this we have asymptotic de Sitter-like behaviors with values of $E=1$ as can be seen the Table (\ref{tabla2}). In the same way, let us note that our Big-rip Type I singularities also occurs with phantom-like behavior with a parameter of state given by Eq. \eqref{weff}, but this type of phantom occur in the context of our total fluid composed of dissipative  DM  and CC. Our results show that it is possible to extend the classification of Big-rip singularity to models where the phantom EoS is effective and not necessarily appears in phantom DE models.

For a positive CC in the early time behavior a remarkable result is that we only have universes without singularities. A special case appears for the solution with $m=1$ which represents scenarios without singularity as we discussed in section \ref{soft}. For this particular solution, beyond not having singularity, is that it's behavior is very similar to the standard model for very small values of viscosity, in addition to being different from other singularity-free models \cite{Softbang1,SoftBang2,Rosen,BigBounce}. This behavior is independent of the sign of the CC. In \cite{Big.Bang} a similar behavior was obtained without the inclusion of the CC. In our solution a CC is considered and the soft Big-bang is characterized by having a zero scale factor at a very past time, 
which is different from the obtained in \cite{Softbang1}.


For a negative CC in the early time behavior a remarkable result is that the singularities only appears in the case with $m=0$, and the constraints in the parameters show that the singularity required values of $\Omega_{\xi}$ that depend on the values of the CC. So, despite the fact that in early times its contribution it is very small, its presence is required for the existence of this singularity. 


Our results indicate that the inclusion of dissipation in the $\Lambda$CDM model leads to solutions where the Big Rip singularities appears without a phantom DE and the avoidance of Big Bang singularities is also possible. Therefore, the dissipation mechanism, which is a more realistic description of cosmic fluid, can alleviate the theoretical problems of phantom DE and initial singularities, and also we can obtain solutions whose behavior is very similar to the standard $\Lambda$CDM model.

\begin{table}[ht]
\begin{center}
\caption{Classification of the early and late times singularities.}
\begin{tabular}{| m{1.4cm} |  m{1.9cm}|m{1.9cm} | m{2.5cm}  | }
\hline\multirow{1}{1cm}[0cm]{\centering\textbf{Solution}}& \multirow{1}{1.7cm}[0cm]{\centering\textbf{Late-Time}}& 
\textbf{Early-Time} &
\multirow{1}{1.7cm}[0cm]{\textbf{Condition}}\\
\hline & \centering{{\textbf{Type 0B (Big-Crunch)}}} \hrule
& \hrule &
\begin{eqnarray*}
\Omega_{\xi}&>&2\\
\Omega_{\Lambda}&=&-\left(\frac{\Omega_{\xi}}{2}\right)^2\\
\end{eqnarray*} \hrule \\
\multirow{2}{1.4cm}[2cm]{\centering\textbf{m=0 and $\Delta_{0}=0$}} & & \multirow{2}{1.9cm}[1.5cm]{\textbf{Type 0A (Big-Bang)}} &
\begin{eqnarray*}
\Omega_{\xi}&<&\frac{8}{3}
\\
\Omega_{\Lambda}&=&-\left(\frac{3\Omega_{\xi}}{8}\right)^2
\end{eqnarray*}
\\ 
\hline &  \centering{{\textbf{Type 0B (Big-Crunch)}}} \hrule& \hrule & 
\begin{eqnarray*}
 \Omega_{\Lambda}<-\left(\frac{\Omega_{\xi}}{2}\right)^2\\
\end{eqnarray*} \hrule \\
\multirow{3}{1.4cm}[3.2cm]{\centering\textbf{m=0 and $\Delta_{0}<0$}} & & \multirow{3}{1.9cm}[1.5cm]{\textbf{Type 0A (Big-Bang)}} &
\begin{eqnarray*}
 \Omega_{\Lambda}<-\left(\frac{3\Omega_{\xi}}{8}\right)^2\\
\end{eqnarray*}
\\ \hline
\multirow{2}{1.4cm}[1.4cm]{\centering\textbf{m=1 and $\Omega_{\Lambda}>0$}} & \multirow{2}{1.9cm}[1.4cm]{\textbf{Type I (Big-Rip)}} & &
\begin{eqnarray*}
\Omega_{\xi}&>&1\\
\gamma_{eff}&<&0
\end{eqnarray*}\\
\hline &
 \centering{{\textbf{Type 0B (Big-Crunch)}}} \hrule& \hrule & 
\begin{eqnarray*}
    \Omega_{\xi}&<& 1\\
\end{eqnarray*}\hrule \\
\multirow{4}{1.4cm}[3.0cm]{\centering\textbf{m=1 and $\Omega_{\Lambda}<0$}} &  \multirow{3}{1.9cm}[1.5cm]{\textbf{Type I (Big-Rip)}} & &
\begin{eqnarray*}
\Omega_{\xi}&>&1\\
\gamma_{eff}&<&0
\end{eqnarray*}
\\ \hline
\end{tabular}
    \label{tabla1}
\end{center}
\end{table}

\begin{table}[H]
\begin{center}
\caption{Classification of the asymptotic behavior for early and late times without singularities.}
\begin{tabular}{| m{1.4cm} |  m{1.9cm}|m{1.9cm} | m{2.8cm}  | }
\hline\multirow{1}{1cm}[0cm]{\centering\textbf{Solution}}& \multirow{1}{1.7cm}[0cm]{\centering\textbf{Late-Time}}& 
\textbf{Early-Time} &
\multirow{1}{1.7cm}[0cm]{\textbf{Condition}}\\
\hline & 
\begin{equation*}
\frac{\frac{\Omega_{\xi}}{\gamma}+\sqrt{\bar{\Delta}_{0}}}{2}
\end{equation*}
\hrule
& \hrule &
\begin{eqnarray*}
&1<\gamma<2\\
&\Omega_{\xi_{0}}>0\\
&-\left(\frac{\Omega_{\xi}}{2\gamma}\right)^2<\Omega_{\Lambda}<0
\end{eqnarray*} \hrule \\
\multirow{2}{1.4cm}[3.5cm]{\centering\textbf{m=0 and $\Delta_{0}>0$}} & & \begin{equation*}
\frac{\frac{\Omega_{\xi}}{\gamma}-\sqrt{\bar{\Delta}_{0}}}{2}
\end{equation*} &
\begin{eqnarray*}
&1<\gamma<2\\
&\Omega_{\xi_{0}}>0\\
&-\left(\frac{\Omega_{\xi}}{2\gamma}\right)^2<\Omega_{\Lambda}<0 \end{eqnarray*}
\\ 
\hline &  \begin{eqnarray*}
E_{ds}=1
\end{eqnarray*} \hrule& \begin{eqnarray*}
E_{ds}=1
\end{eqnarray*} \hrule & 
\begin{eqnarray*}
\Omega_{\xi}=2\gamma\\
\Omega_{\Lambda}=-1
\end{eqnarray*} \hrule \\
\multirow{3}{1.4cm}[3cm]{\centering\textbf{m=0 and $\Delta_{0}=0$}} & \multirow{3}{1.4cm}[1.5cm]{\centering\textbf{$E_{ds}=\frac{\Omega_{\xi}}{2}$}} &  &
\begin{eqnarray*}
\Omega_{\xi}&<&2\\
\Omega_{\Lambda}&=&-\left(\frac{\Omega_{\xi}}{2}\right)^2
\end{eqnarray*}
\\ \hline & \begin{eqnarray*}
E_{ds}=1
\end{eqnarray*} \hrule& \begin{eqnarray*}
E_{ds}=1
\end{eqnarray*} \hrule & 
\begin{eqnarray*}
\Omega_{\xi}=1
\end{eqnarray*} \hrule \\
& \begin{eqnarray*}
E_{ds}=\sqrt{\Omega_{\Lambda}}
\end{eqnarray*} \hrule& \hrule & 
\begin{eqnarray*}
\Omega_{\xi}<1
\end{eqnarray*} \hrule \\
\multirow{4}{1.4cm}[3.5cm]{\centering\textbf{m=1 and $\Omega_{\Lambda}>0$}} &  &\multirow{3}{1.9cm}[1.5cm]{\textbf{Soft-Big-Bang \\ $E_{ds}=\frac{4}{3\Omega_{\xi}}$}}  &
\begin{eqnarray*}
\Omega_{\xi}<\frac{4}{3}
\end{eqnarray*}
\\ \hline
 & \begin{eqnarray*}
E_{ds}=1
\end{eqnarray*} \hrule& \begin{eqnarray*}
E_{ds}=1
\end{eqnarray*} \hrule & 
\begin{eqnarray*}
\Omega_{\xi}=1
\end{eqnarray*} \hrule \\
\multirow{4}{1.4cm}[2.1cm]{\centering\textbf{m=1 and $\Omega_{\Lambda}<0$}} &  &\multirow{3}{1.9cm}[1.5cm]{\textbf{Soft-Big-Bang \\ $E_{ds}=\frac{4}{3\Omega_{\xi}}$}}  &
\begin{eqnarray*}
\Omega_{\xi}<\frac{4}{3}
\end{eqnarray*}
\\ \hline
\end{tabular}
    \label{tabla2}
\end{center}
\end{table}


\section*{Acknowledgments}

Norman Cruz acknowledges the support of Universidad de Santiago de Chile (USACH), through Proyecto DICYT N$^{\circ}$ 042131CM, Vicerrectoría de Investigación, Desarrollo e Innovación. Esteban González acknowledges the support of Dirección de Investigación y Postgrado at Universidad de Aconcagua. Jose Jovel acknowledges ANID-PFCHA/Doctorado Nacional/2018-21181327.

\vspace{190mm}
\bibliography{JCAP.bib}

\end{document}